\pdfoutput=1
\documentclass[manuscript]{acmart}
\usepackage[ruled,linesnumbered]{algorithm2e}
\usepackage{multirow}
\usepackage{subfigure}
\AtBeginDocument{%
  \providecommand\BibTeX{{%
    \normalfont B\kern-0.5em{\scshape i\kern-0.25em b}\kern-0.8em\TeX}}}

\setcopyright{acmcopyright}
\copyrightyear{2018}
\acmYear{2018}
\acmDOI{XXXXXXX.XXXXXXX}

\acmJournal{JACM}
\acmVolume{37}
\acmNumber{4}
\acmArticle{111}
\acmMonth{8}

\begin{document}

%%
%% The "title" command has an optional parameter,
%% allowing the author to define a "short title" to be used in page headers.
\title{\algo~: You can Backdoor Personalized Federated Learning}

%%
%% The "author" command and its associated commands are used to define
%% the authors and their affiliations.
%% Of note is the shared affiliation of the first two authors, and the
%% "authornote" and "authornotemark" commands
%% used to denote shared contribution to the research.
\author{Tiandi Ye}
\email{52205903002@stu.ecnu.edu.cn}
\orcid{0000-0003-0169-457X}
\affiliation{
  \institution{East China Normal University}
  \city{Shanghai}
  \country{China}
  \postcode{200062}
}

\author{Cen Chen}
\email{cenchen@dase.ecnu.edu.cn}
\orcid{0000-0003-0325-1705}
\affiliation{
  \institution{East China Normal University}
  \city{Shanghai}
  \country{China}
  \postcode{200062}
}
\authornote{Corresponding author.}

\author{Yinggui Wang}
\affiliation{
  \institution{Ant Group}
  \city{Beijing}
  \country{China}
  \postcode{100020}
}
\email{wyinggui@gmail.com}
\orcid{0000-0002-6686-6603}

\author{Xiang Li}
\email{xiangli@dase.ecnu.edu.cn}
\orcid{0000-0003-0945-145X}
\affiliation{
  \institution{East China Normal University}
  \city{Shanghai}
  \country{China}
  \postcode{200062}
}

\author{Ming Gao}
\email{mgao@dase.ecnu.edu.cn}
\orcid{0000-0002-5603-2680}
\affiliation{
  \institution{East China Normal University}
  \city{Shanghai}
  \country{China}
  \postcode{200062}
}

%%
%% By default, the full list of authors will be used in the page
%% headers. Often, this list is too long, and will overlap
%% other information printed in the page headers. This command allows
%% the author to define a more concise list
%% of authors' names for this purpose.
\newcommand{\ytd}[1]{\textcolor{red}{[#1 --ytd]}}
\newcommand{\cc}[1]{{\color{brown}[#1 --cc]}}
\newcommand{\li}[1]{{\color{blue}[#1 --lx]}}
\newcommand{\algo}[1]{{BapFL}}
\newcommand{\basealgo}[1]{{BapFL$^{-}$}}
\renewcommand{\shortauthors}{Tiandi Ye and Cen Chen, et al.}

%%
%% The abstract is a short summary of the work to be presented in the
%% article.
\begin{abstract}
In federated learning (FL), 
malicious clients could manipulate the predictions of the trained model through backdoor attacks, 
posing a significant threat to the security of FL systems. 
Existing research primarily focuses on backdoor attacks and defenses 
within the generic federated learning scenario, where all clients collaborate to train a single global model. 
A recent study conducted by \citeauthor{qin2023revisiting} (\citeyear{qin2023revisiting}) 
marks the initial exploration of backdoor attacks 
within the personalized federated learning (pFL) scenario, 
where each client constructs a personalized model based on its local data. 
Notably, the study demonstrates that pFL methods with \textit{parameter decoupling} can significantly enhance robustness against backdoor attacks. 
However, in this paper, we whistleblow that 
pFL methods with parameter decoupling are still vulnerable to backdoor attacks. 
The resistance of pFL methods with parameter decoupling is attributed to the heterogeneous classifiers between malicious clients and benign counterparts. 
We analyze two direct causes of the heterogeneous classifiers: 
(1) data heterogeneity inherently exists among clients and 
(2) poisoning by malicious clients further exacerbates the data heterogeneity. 
To address these issues, 
we propose a two-pronged attack method, 
\algo~, 
which comprises two simple yet effective strategies: 
(1) poisoning only the feature encoder while keeping the classifier fixed and 
(2) diversifying the classifier through noise introduction to simulate that of the benign clients. 
Extensive experiments on three benchmark datasets under varying conditions demonstrate the effectiveness of our proposed attack. 
Additionally, we evaluate the effectiveness of six widely used defense methods and find that \algo~ still poses a significant threat even in the presence of the best defense, Multi-Krum. 
We hope to inspire further research on attack and defense strategies in pFL scenarios. The code is available at: https://github.com/BapFL/code.

\end{abstract}

%%
%% The code below is generated by the tool at http://dl.acm.org/ccs.cfm.
%% Please copy and paste the code instead of the example below.
%%


%%
%% Keywords. The author(s) should pick words that accurately describe
%% the work being presented. Separate the keywords with commas.

\received{20 February 2007}
\received[revised]{12 March 2009}
\received[accepted]{5 June 2009}

%%
%% This command processes the author and affiliation and title
%% information and builds the first part of the formatted document.
\maketitle

\section{Introduction}
\label{sec:intro}
Federated learning (FL) has gained significant popularity 
due to its ability to enable collaborative model training among 
multiple clients while preserving their data privacy~\cite{kairouz2021advances,li2020federated}. 
In FL, FedAvg~\cite{mcmahan2017communication} stands as the de facto FL method, which updates the server model by aggregating the model parameters from the clients. 
However, the performance of the model can be significantly affected by the non-IID (non-independent and non-identically distributed) nature of the data. 
To address this challenge, \textit{personalized federated learning} (pFL) has emerged as a promising solution, 
as it allows personalized models for each participating client. 
Existing pFL methods can be broadly categorized based on the strategies employed to achieve personalization. 
These strategies include 
parameter decoupling~\cite{arivazhagan2019federated,collins2021exploiting,oh2021fedbabu,chen2021bridging,li2021fedbn,chen2022learning}, 
regularization~\cite{li2021ditto,t2020personalized}, 
clustering~\cite{sattler2020clustered,ghosh2020efficient,mansour2020three,ye2023robust}, and model interpolation~\cite{mansour2020three}.

The distributed training nature of FL makes it particularly vulnerable to backdoor attacks~\cite{bagdasaryan2020backdoor, sun2019can, qin2023revisiting, chen2020backdoor, zhuang2023backdoor, zhang2022neurotoxin}, which manipulate the target model's behavior by implanting specific triggers~\cite{gu2019badnets}. 
Recently, 
\citet{qin2023revisiting} demonstrated that 
pFL methods with parameter decoupling (i.e., partial model-sharing) 
can effectively alleviate backdoor attacks under the black-box assumption. 
There are two primary types of parameter decoupling. 
The first decomposes the model into the feature encoder and the classifier, 
as exemplified by 
FedPer~\cite{arivazhagan2019federated}, 
FedRep~\cite{collins2021exploiting}, 
FedRod~\cite{chen2021bridging}, 
FedBABU~\cite{oh2021fedbabu} 
and FedMC~\cite{chen2022learning}. 
In these methods, 
only the feature encoder is shared with the server, while the classifier remains locally for personalization. 
And the second retains the Batch Normalization layers locally and synchronizes only the remaining parameters with the server, as demonstrated by FedBN~\cite{li2021fedbn}.

\begin{figure*}[htbp!]
    \centering
    \includegraphics[width=0.5\linewidth,trim=0 0 0 0,clip]{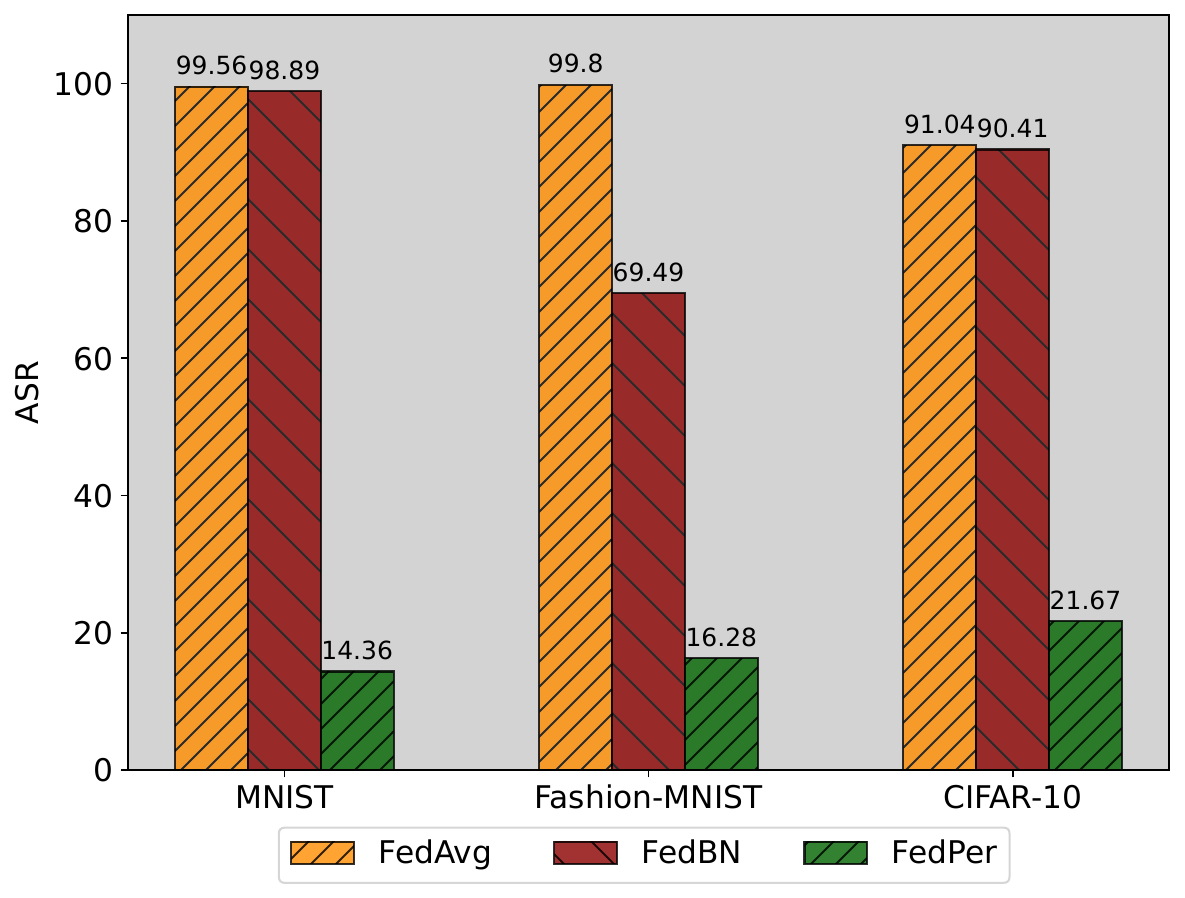}
    \caption{
    Attack success rate (ASR) of black-box backdoor attacks against FedAvg and two pFL methods based on parameter decoupling, namely FedBN and FedPer. 
    Details of the attack setup are discussed in Section~\ref{sec:exp-setup}.}
    \label{fig:why-target-fedper}
\end{figure*}

In this paper, we whistleblow that pFL methods with parameter decoupling are still vulnerable to backdoor attacks. 
Specifically, in Figure~\ref{fig:why-target-fedper}, we conduct black-box backdoor attacks against FedAvg, FedBN, and FedPer on three different datasets. 
It is evident that FedAvg exhibits a high vulnerability to such attacks, whereas FedBN provides negligible defense. 
In contrast, FedPer exhibits remarkable resistance to backdoor attacks. 
The remarkable resistance can be attributed to the significant divergence in local classifiers among clients. 
As a result, malicious clients are effectively impeded from generalizing the success of their attacks to the benign counterparts. 

%%%%%%%

\begin{figure*}[htbp!]
    \centering
    \includegraphics[width=0.5\linewidth,trim=0 0 0 0,clip]{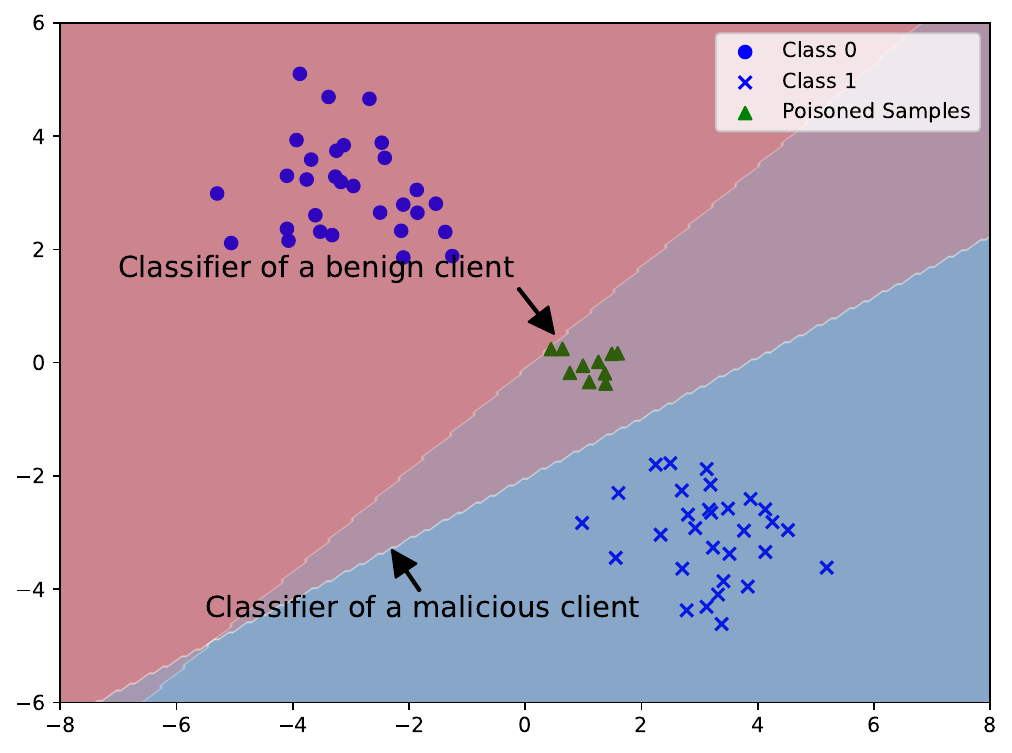}
    \caption{
    {
    Illustration of factor $F_{2}$. 
    Green triangles denote the poisoned samples whose ground truth labels are class 1, and the target label is class 0. 
    Poisoning has significantly altered the data distribution, resulting in a noticeable shift in the classifier boundary compared to that of the benign client.}
    }
    \label{fig:illA}
\end{figure*}

Classifier heterogeneity arises from two primary factors: 
($F_{1}$) data heterogeneity, such as class imbalance and concept-shift, inherently exists among clients, and 
($F_{2}$) poisoning by malicious clients exacerbates the distribution disparities between malicious and benign clients. 
As illustrated in Figure~\ref{fig:illA}, 
after poisoning, the data distribution changes significantly in terms of class distribution $p(y)$ and conditional distribution $p(x|y)$. 
The process of executing a backdoor attack requires the malicious client to adjust the classifier in order to classify the poisoned samples (originally belonging to the ground truth class 1) as class 0. 
Consequently, this manipulation leads to a distinct classifier boundary compared to that of the benign client. 
Building upon this observation, we propose a strategy for conducting backdoor attacks by targeting the feature representation of the poisoned samples towards the desired target class while keeping the classifier fixed. 

\begin{figure*}[htbp!]
    \centering
    \includegraphics[width=0.5\linewidth,trim=0 0 0 0,clip]{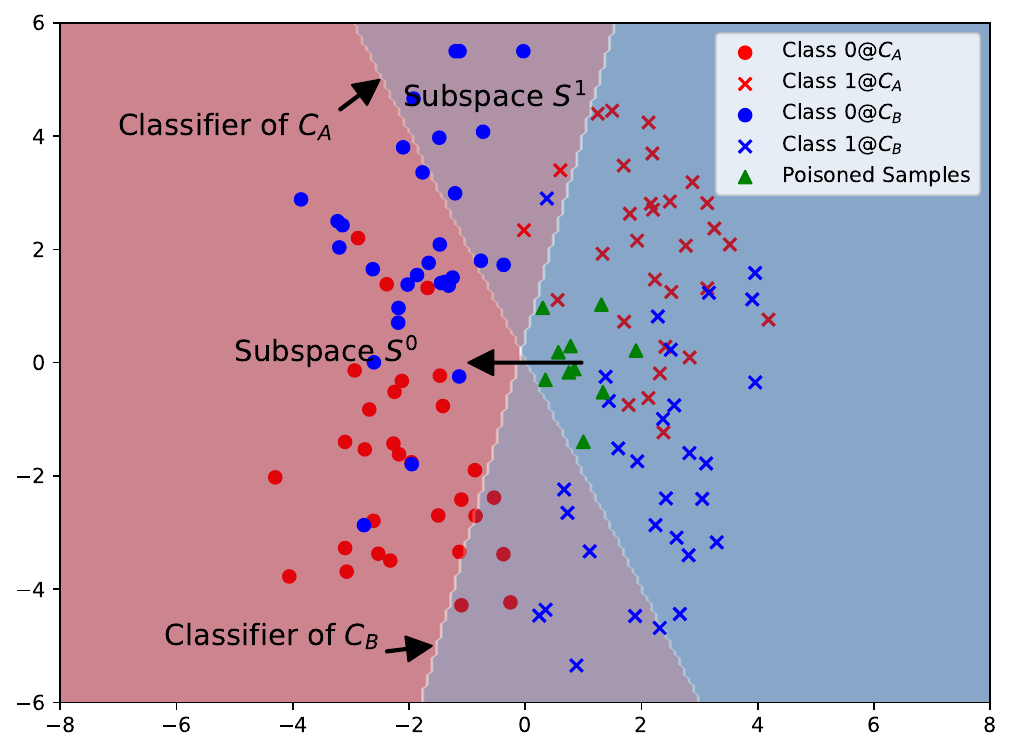}
    \caption{
    {
    Illustration of factor $F_{1}$. 
    The presence of data heterogeneity between client $C_{A}$ and client $C_{B}$ results in significant variation in their respective classifiers.}
    }
    \label{fig:illB}
\end{figure*}

Furthermore, as depicted in Figure~\ref{fig:illB}, the presence of data heterogeneity among clients results in significant variation in their respective classifiers. 
From the perspective of the malicious client $C_{B}$, steering the representation of the poisoned samples towards subspace $S^{1}$ may appear advantageous. 
However, even if the representation falls within $S^{1}$, it continues to be correctly categorized as class 1 by the benign client $C_{A}$. 
With a broader perspective, to effectively target both client $C_{A}$ and $C_{B}$, it is preferable to guide the representation of the poisoned samples towards subspace $S^{0}$. 
Building upon this insight, we propose to enhance the attack's 
generalizability by introducing noise to diversify the classifiers. The goal is to align the representation of poisoned samples as closely as possible with that of the target class for all clients involved in the federated learning process. 

By combining the aforementioned two strategies, i.e.,
(1) solely poisoning the feature encoder and 
(2) enhancing classifier diversity through noise introduction, 
we design a practical and easy-to-implement backdoor attack, 
called \algo~. 
Our attack is designed to operate under the white-box assumption, where the attacker has full access to the local training process. 
Subsequently, we conduct extensive experiments on three benchmark datasets. 
Results show that our proposed \algo~ achieved an impressive attack success rate (ASR) of 94.21\% on MNIST, 80.93\% on Fashion-MNIST, and 58.89\% on CIFAR-10, demonstating that the decoupling strategy  gives a false sense of security. 
Furthermore, we investigate the effectiveness of six widely used defense methods, including 
Gradient Norm-Clipping~\cite{sun2019can}, 
Median~\cite{yin2018byzantine}, 
Trimmed Mean~\cite{yin2018byzantine}, 
Multi-Krum~\cite{blanchard2017machine}, 
Fine-Tuning~\cite{qin2023revisiting}, 
and Simple-Tuning~\cite{qin2023revisiting}. 
We find that only Multi-Krum provides effective defense. 
However, with the projected gradient descent (PGD)~\cite{wang2020attack}, 
\algo~ is still able to bypass the defense of Multi-Krum
and achieves an ASR of 70.60\%. 

We summarize our contributions as follows:
\begin{itemize}
  \item We alert to the FL community that 
  even pFL methods with parameter decoupling are still highly vulnerable to backdoor attacks. 
  We introduce a practical and easy-to-implement backdoor attack, namely \algo~, {which achieves remarkable attack performance while incurring negligible computational and storage costs}.
  \item We conduct extensive experiments on three benchmark datasets to thoroughly evaluate the effectiveness of \algo~. 
  The experimental results validate its efficacy, and further experiments demonstrate its superiority over baselines under varying conditions, including data heterogeneity, feature encoder size, number of malicious attackers, and attack frequency. 
  An additional ablation study further demonstrates the effectiveness of our introduced strategies.
  \item Additionally, we {assess the effectiveness of six widely used} defense methods 
  and find that even in the presence of these defenses, \algo~ can still effectively target personalized models, yielding an ASR of 70.60\%. 
  We hope our findings could stimulate further research on both attack and defense strategies within the context of pFL scenarios.
\end{itemize}

\section{Related Work}
\subsection{Personalized Federated Learning}
FL is a distributed machine learning paradigm, 
which enables training models on decentralized data sources 
and ensuring data privacy. 
We consider an FL system with $N$ clients, 
where each client has a private dataset 
$\mathcal{D}_{i}=\{(x_{j}, y_{j})\}_{j=1}^{|\mathcal{D}_{i}|}$. 
The optimization objective of the generic FL can be formulated as 
\begin{equation}
\begin{gathered}
\underset{\theta}{\text{min}} 
~\sum_{i=1}^{N} \frac{|\mathcal{D}_{i}|}{|\mathcal{D}|}\mathcal{L}_{i}(\theta), \\ 
\text{where} ~\mathcal{L}_{i} = \frac{1}{|\mathcal{D}_{i}|} \sum_{j=1}\ell(x_{j}, y_{j}; \theta). 
\end{gathered}
\label{eq:fl-obj}
\end{equation}
Here, $\mathcal{D}=\cup_{i}\mathcal{D}_{i}$ is the aggregated dataset from all clients, 
$\theta$ is the global parameters, 
$\mathcal{L}_{i}(\theta)$ is the empirical risk of client $c_{i}$, 
and $\ell$ refers to the loss function applied to each data instance, 
commonly the cross-entropy loss for classification tasks. 
FedAvg~\cite{mcmahan2017communication} is the first proposed solution to Equation\ref{eq:fl-obj}, which iterates between local training and global aggregation for multiple rounds of communication. 
Besides FedAvg, other generic FL methods, such as FedProx\cite{li2020federated} and SCAFFOLD~\cite{karimireddy2020scaffold}, output a universal model for all clients. 

In terms of the test accuracy of each client, pFL proves to be a more effective learning paradigm, especially in the presence of data heterogeneity. 
Existing pFL methods can be broadly categorized based on the strategies employed to achieve personalization, e.g., 
parameter decoupling~\cite{arivazhagan2019federated,collins2021exploiting,oh2021fedbabu,chen2021bridging,li2021fedbn,chen2022learning}, 
regularization~\cite{li2021ditto,t2020personalized}, 
clustering~\cite{sattler2020clustered,ghosh2020efficient,mansour2020three,ye2023robust}, and model interpolation~\cite{mansour2020three}. 
For a detailed discussion, we refer readers to the survey~\cite{tan2022towards}. 
Here, we primarily focus on methods based on parameter decoupling, which involves decomposing the model into global and local components. There are two primary types of parameter decoupling.
The first type keeps the Batch Normalization~\cite{ioffe2015batch} layers local and synchronizes only the remaining parameters with the server, as demonstrated by FedBN~\cite{li2021fedbn}.
Another predominant type divides the model into the feature encoder $\omega$ and the classifier $\psi$~\cite{arivazhagan2019federated, collins2021exploiting, oh2021fedbabu, chen2021bridging, chen2022learning}.
The optimization objective of such methods can be expressed as follows: 
\begin{equation}
\begin{gathered}
\underset{\{\omega, \psi_{1}, \cdots, \psi_{N}\}}{\text{min}} 
~\sum_{i=1}^{N} \frac{|\mathcal{D}_{i}|}{|\mathcal{D}|}\mathcal{L}_{i}(\omega, \psi_{i}), \\ 
\text{where} ~\mathcal{L}_{i} = \frac{1}{|\mathcal{D}_{i}|} \sum_{j=1}\ell(x_{j}, y_{j}; \omega, \psi_{i}). 
\end{gathered}
\end{equation}
Here, $\mathcal{L}_{i}(\omega, \psi_{i})$ is the empirical risk of client $i$. 
FedPer~\cite{arivazhagan2019federated} and FedRep~\cite{collins2021exploiting} learn a global feature encoder across clients 
and personalized classifiers for each client. 
In FedPer~\cite{arivazhagan2019federated}, 
each client jointly updates the feature encoder and the classifier, 
whereas in FedRep~\cite{collins2021exploiting}, 
each client sequentailly updates the classifier and the feature encoder. 
In FedBABU~\cite{oh2021fedbabu}, each client learns the feature encoder with the randomly initialized classifier and obtains the personalized model by fine-tuning the global model. 
FedRod~\cite{chen2021bridging} employs a one-body-two-head architecture comprising a feature encoder and two classifiers (heads). The personalized classifier is trained locally and never shared with the server. 
In contrast, FedMC~\cite{chen2022learning} utilizes a two-body-one-head architecture that includes a personalized classifier, a personalized feature encoder, and a global feature encoder. 
The personalized components are maintained locally.

\subsection{Backdoor Attack and Defense}
Recently, 
backdoor attacks have gained significant attention 
due to their potential to compromise the security and trustworthiness of deep learning models. 
In a backdoor attack, 
malicious clients implant hidden triggers into the target model. 
During inference, any input containing the trigger will be predicted as the predefined target label regardless of its ground truth, 
while ensuring that the model functions normally on clean inputs~\cite{li2022backdoor, gu2019badnets}. 

Due to the decentralized nature of FL, 
malicious clients may manipulate their local models 
and inject backdoors into the global model during training~\cite{bagdasaryan2020backdoor, wang2020attack, zhang2022flip, qin2023revisiting, chen2020backdoor, zhang2022neurotoxin, xie2020dba}. 
Backdoor attacks in FL 
can be broadly categorized into two categories: 
white-box attacks and black-box attacks~\cite{lyu2020threats}. 
In white-box attacks, 
malicious clients have full access to the local training process, while in black-box attacks, adversarial clients are limited to manipulating the local datasets. 
Scaling attack~\cite{bagdasaryan2020backdoor} scales the model parameters before synchronizing with the server 
to cancel out the contributions from the benign clients 
and implement model replacement. 
Under the assumption that multiple malicious clients collaborate and collude, DBA~\cite{xie2020dba} launch a distributed-trigger backdoor attack, which decomposes the global trigger into multiple parts and distributes them among multiple attackers. 
Experiment results have demonstrated that DBA outperforms the single-trigger approach in terms of both attack effectiveness and stealth. 
Neurotoxin~\cite{zhang2022neurotoxin} extends the duration and persistence of backdoor attacks by projecting the adversarial gradient onto the subspace that remains unused by benign users during retraining. 

To defend against backdoor attacks, many strategies have been proposed, such as 
Gradient Norm-Clipping~\cite{sun2019can}, Median~\cite{yin2018byzantine}, 
Trimmed Mean~\cite{yin2018byzantine}, 
Krum and Multi-Krum~\cite{blanchard2017machine}. 
Recent work~\cite{qin2023revisiting} conducts the first study of backdoor attacks in the pFL framework 
and shows that 
pFL methods with partial model-sharing 
can effectively alleviate black-box backdoor attacks. 
Inspired by the defense mechanism of pFL methods of partial model-sharing, they provide a defense method that re-initializes the private parameters and retrains them on clean samples with the global parameters fixed.

\section{Methodology}
\subsection{Threat Model} 
In this paper, 
we backdoor attack against pFL methods 
under the white-box assumption, 
where malicious clients can poison local data and manipulate the local training procedure. 
Similar to previous studies~\cite{xie2020dba,qin2023revisiting},
we focus on the image classification task, which has been widely used in research on backdoor attacks. 

\subsubsection{Attacker's Capacities}
We assume that 
attackers have full access to their local training. 
They can poison a subset of local data and modify the training components, such as training loss and training schedule. 

\subsubsection{Attacker's Goals}
Malicious clients aim to implant backdoors into the global featuer encoder. 
During testing on images that contain the trigger, the predictions of the personalized models will be arbitrarily altered to the target class. 
In addition, a successful attack should ensure stealthiness, i.e., the performance of the personalized model on clean samples should not be significantly reduced. 

\subsubsection{Targeted pFL Methods}
Parameter-decoupling based pFL methods, such as FedPer, FedRep, FedRod, FedBABU, and FedMC, all share a common principle: decomposing the model into a global feature encoder and a personalized classifier. In this paper, our primary focus is on attacking the widely adopted parameter decoupling strategy. 
Without explicit declaration, we specifically target FedPer due to its simplicity in the training process. 
But it's worth noting that the analysis can be easily extended to other methods.

\subsection{\algo~: Backdoor Attack against pFL} 
The resistance of parameter-decoupling based pFL methods 
against backdoor attacks 
lies in the significant divergence in classifiers between the malicious and benign clients. 
To overcome the defense, we design a two-pronged attack method called \algo~. 
This attack encompasses two crucial poisoning strategies: 
\begin{itemize}
  \item $\text{PS}_{1}$: Poison only the feature encoder while keeping the classifier fixed. 
  This strategy allows us to circumvent factor $F_{2}$, as analyzed in Section~\ref{sec:intro}, since the classifier is never exposed to the poisoned samples. 
  \item $\text{PS}_{2}$: Diversify the classifier by introducing noise to simulate that of the benign clients. 
  This strategy primarily targets addressing factor $F_{1}$. 
\end{itemize}
We have detailed \algo~ in Algorithm~\ref{alg:algorithm}. 
Next, we will delve into these two strategies individually.

\subsubsection{Poisoning Only Feature Encoder}
Generalizing the attack to benign clients 
entails aligning the classifier of malicious clients with that of benign ones. 
To achieve this, 
we first apply strategy $S_1$ to circumvent factor $F_{2}$. 
Specifically, during each local iteration, 
malicious client $C_{a}$ poisons $b$ samples within a batch (line 13). 
Subsequently, it updates the entire model to minimize the model's loss on clean samples (line 14-15). 
Afterwards, it proceeds to update only the feature encoder to minimize the model's loss on the poisoned samples while keeping the classifier unchanged (line 18). 
The optimization of the malicious client's classifier aligns with that of benign clients, utilizing only knowledge from clean samples.

\subsubsection{Diversifying Local Classifier}
We formulate the optimization objective of \algo~ as follows: 
\begin{equation}
\begin{gathered}
\underset{\omega}{\text{min}} 
~\frac{1}{N} \sum_{i=1}^{N} \mathcal{\Tilde{L}}_{i}(\omega, \psi_{i}). 
\end{gathered}
\label{eq:origin-attacker-objective}
\end{equation}
Here, we use $\mathcal{\Tilde{L}}_{i}(\omega, \psi_{i})$ to represent the loss of the model parameterized by ($\omega$, $\psi_{i}$) on the corresponding poisoned dataset $\mathcal{\Tilde{D}}_{i}$. 
However, directly optimizing Equation~\ref{eq:origin-attacker-objective} is infeasible, as the attacker (denoted as $C_{a}$) lacks access to other clients' classifiers $\{\psi_{i}\}_{i=1}^{N\setminus a}$. 

A straightforward strategy is simulating other clients' classifiers with randomly initialized parameters. 
However, a randomly generated classifier is likely to significantly deviate from that of benign clients. 
And we have empirically observed that randomly generated classifiers lead to unstable and slow convergence in terms of attack. 
Therefore, 
we suggest introducing noise to the attacker's classifier $\psi_{a}$ 
to mimic that of benign clients. 
Specifically, 
we sample noise from an isotropic multivariate Gaussian distribution $\mathcal{N}(0, \sigma)$ 
and then add it to $\psi_{a}$. 
The optimization objective in Equation~\ref{eq:origin-attacker-objective} is reformulated as follows: 
\begin{equation}
\begin{gathered}
\underset{\omega}{\text{min}} 
~\frac{1}{N} \sum_{i=1}^{N} \mathcal{\Tilde{L}}_{i}(\omega, \psi_{a}+\epsilon_{i}), \text{where} ~\epsilon_{i} \sim \mathcal{N}(0, \sigma).
\end{gathered}
\label{eq:reformulated-attacker-objective-with-noise}
\end{equation}
The main objective of backdoor attacks is to compel the model to learn the mapping from the trigger pattern to the target label. 
Consequently, we can approximately substitute 
$\mathcal{\Tilde{L}}_{i}(\omega, \psi_{a}+\epsilon_{i})$ with $\mathcal{\Tilde{L}}_{a}(\omega, \psi_{a}+\epsilon_{i})$. 
As a result, we can simplify the optimization objective in Equation~\ref{eq:reformulated-attacker-objective-with-noise} as follows: 
\begin{equation}
\begin{gathered}
\underset{\omega}{\text{min}} 
~\frac{1}{N} \sum_{i=1}^{N} \mathcal{\Tilde{L}}_{a}(\omega, \psi_{a}+\epsilon_{i}), \text{where} ~\epsilon_{i} \sim \mathcal{N}(0, \sigma), 
\end{gathered}
\label{eq:simplified-attacker-objective}
\end{equation}
which can be optimized by the attacker itself without requiring any information from other clients. 
In FL, clients typically perform local training for multiple iterations. 
To enhance efficiency, we introduce noise $\epsilon$ into $\psi_{a}$ once during each iteration (line 16-17). 
By consistently sampling and injecting noise, the attacker gains exposure to a diverse set of classifiers, thereby making it possible to attack other clients with different data distributions. 
By combining the above two strategies, $\text{PS}_{1}$ and $\text{PS}_{2}$, we obtain \algo~. Notably, in comparison to the black-box attack, \algo~ introduces almost no computational and storage overhead. 

In addition, in \algo~, we randomly sample $\epsilon$ from $\mathcal{N}(0, \sigma)$. 
For further efficiency, 
one can formulate the local objective 
$\mathcal{\Tilde{L}}_{a}(\omega, \psi_{a}+\epsilon)$ 
as a minimax problem 
$
\underset{\omega}{\text{min}} 
~\underset{\epsilon}{\text{max}}~\mathcal{\Tilde{L}}_{a}(\omega, \psi_{a}+\epsilon)
$ and solve it with alternative optimization. 
We leave it for future work. 

\begin{algorithm}[htbp!]
\SetAlgoLined
\caption{Training of \algo~}
\label{alg:algorithm}
\textbf{Input}: initial parameters of the feature encoder $\omega^{0}$ and local classifiers $\{\psi_{i}\}_{i=1}^{N}$, 
total communication rounds $R$, 
local iterations $\tau$, 
local minibatch size $B$, 
learning rate $\eta$, 
variance of an isotropic Gaussian distribution $\sigma$, 
number of participating clients per round $M$, and 
number of poisoned samples in a batch $b$. \\
\textbf{Output}: $\omega^{R}$ and $\{\psi_{i}\}_{i=1}^{N}$ \\
\textbf{[Training]} \\
\For{{each round} $r$ : 0 {to} $R-1$} {
    Select a set of clients $\mathcal{S}_{r}$ of size $M$ \\
    \For{{each client} $c_{i} \in \mathcal{S}_{r}$} {
        // Local Training \\
        $\omega_{i} \leftarrow \omega^{r}$ \\
        $\mathcal{B}_{i} \leftarrow$ Split local dataset $\mathcal{D}_i$ into batches of size $B$ \\
        \For{{each iteration} $t$ : 1 {to} $\tau$} {
            Sample a batch $(X_{t}, Y_{t}) \sim \mathcal{B}_{i}$ \\
            \If{$c_{i}$ is an attacker} {
                $(X_{t,\text{clean}}, Y_{t,\text{clean}}), (X_{t,\text{poison}}, Y_{t,\text{poison}}) \leftarrow$ Randomly poison $b$ samples in the batch $(X_{t}, Y_{t})$\\
                $\omega_{i} \leftarrow \omega_{i} - \eta \nabla_{\omega_{i}} \mathcal{L}_{i}(X_{t,\text{clean}}, Y_{t,\text{clean}})$ \\
                $\psi_{i} \leftarrow \psi_{i} - \eta \nabla_{\psi_{i}} \mathcal{L}_{i}(X_{t,\text{clean}}, Y_{t,\text{clean}})$ \\
                $\epsilon \leftarrow$ Sample a noise from $\mathcal{N}(0,\sigma)$ \\
                $\psi_{i} \leftarrow \psi_{i} + \epsilon$ \\ 
                $\omega_{i} \leftarrow \omega_{i} - \eta \nabla_{\omega_{i}} \mathcal{L}_{i}(X_{t,\text{poison}}, Y_{t,\text{poison}})$ \\ 
            } \Else {
                $\omega_{i} \leftarrow \omega_{i} - \eta \nabla_{\omega_{i}} \mathcal{L}_{i}(X_{t}, Y_{t})$ \\
                $\psi_{i} \leftarrow \psi_{i} - \eta \nabla_{\psi_{i}} \mathcal{L}_{i}(X_{t}, Y_{t})$ \\
            }
        }
        $\omega_{i}^{r+1} \leftarrow \omega_{i}$ \\
        Client $c_{i}$ sends $\omega_{i}^{r+1}$ back to server \\
    }
    // Server Aggregation \\
    $\omega^{r+1} \leftarrow \sum_{c_{i} \in \mathcal{S}_{r}} \frac{|\mathcal{D}_{i}|}{\sum_{c_{i} \in \mathcal{S}_{r}}|\mathcal{D}_{i}|} \omega_{i}^{r+1}$ \\
}
\end{algorithm}

\begin{figure}[htbp!]
    \centering
    \includegraphics[width=0.5\linewidth,trim=0 0 0 0,clip]{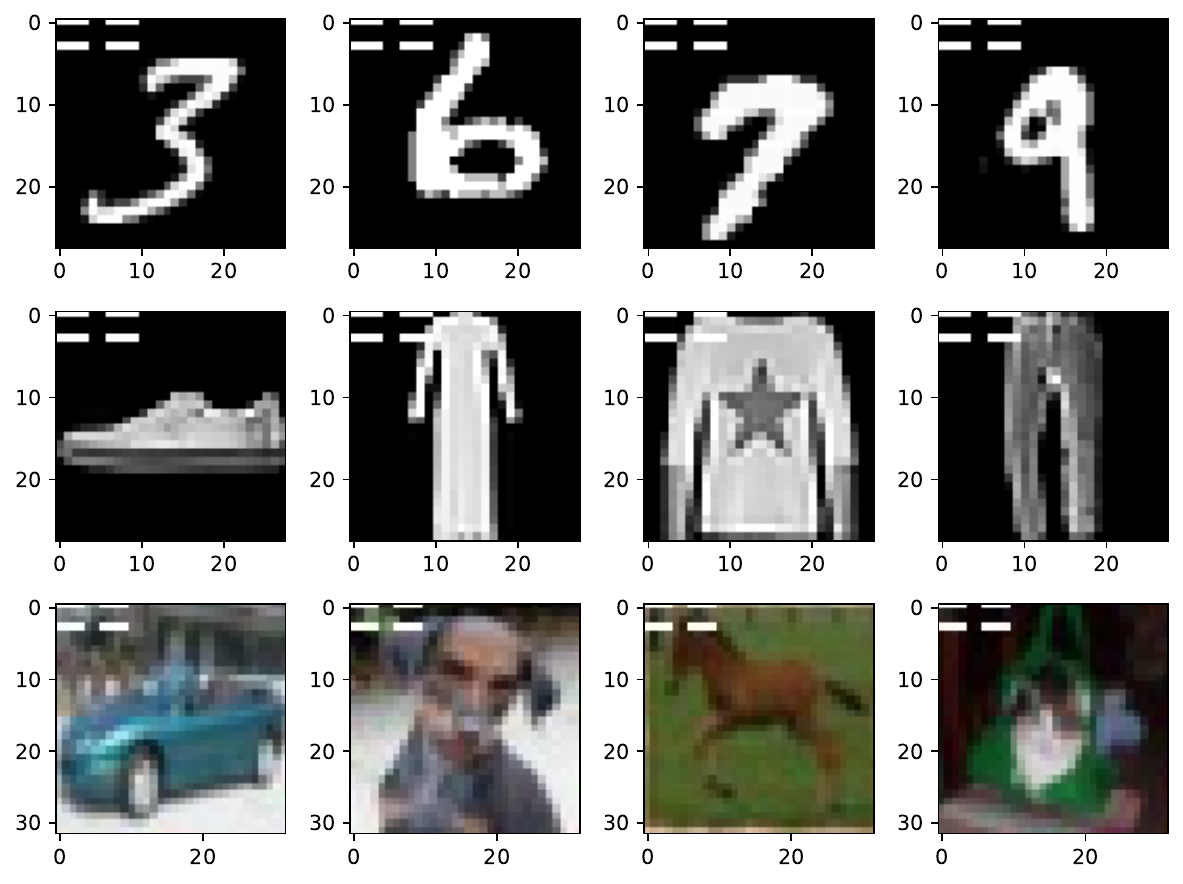}
    \caption{
    Poisoned inputs with a grid pattern trigger from MNIST (top row), Fashion-MNIST (medium row), and CIFAR-10 (bottom row).}
    \label{fig:triggers}
\end{figure}

\section{Attack Evaluation}
In this section, 
we conduct experiments on multiple benchmark datasets to evaluate the attack performance of \algo~ through
answering the following research questions: 
\begin{itemize}
  \item RQ1 (Section~\ref{sec:rq1}): Does \algo~ outperform other backdoor attacks in the pFL setting? 
  \item RQ2 (Section~\ref{sec:rq2}): How does \algo~ perform under varying levels of data heterogeneity? 
  Can \algo~ maintain its effectiveness in strongly heterogeneous settings? 
  \item RQ3 (Section~\ref{sec:rq3}): How does the number of sharing layers impact the attack performance of \algo~ and the baseline attacks? 
  \item RQ4 (Section~\ref{sec:rq4}): Intuitively, backdoor attacks tend to perform better with more malicious clients; can \algo~ effectively launch attacks with only one malicious client? 
  \item RQ5 (Section~\ref{sec:rq5}): Frequent attacks might lack stealth; how does \algo~ perform with varying attack intervals? 
  \item RQ6 (Section~\ref{sec:rq6}): Do strategies $S_1$ and $S_2$ indeed contribute to the improvement in \algo~'s attack performance? 
\end{itemize}

\subsection{Experiment Setup}
\label{sec:exp-setup}
\subsubsection{Datasets and Models} 
We conduct experiments on three widely-used datasets: 
MNIST~\cite{lecun1998gradient}, 
Fashion-MNIST~\cite{xiao2017fashion}, 
and CIFAR-10~\cite{cifar10}. 
Data heterogeneity might stem from many factors, such as class imbalance and concept shift. 
Given that class imbalance is more likely to lead to heterogeneous classifiers, 
in this paper, 
we specifically focus on the class imbalance setting. To simulate heterogeneous environments with class imbalance, 
we generate 50 clients using Dirichlet allocation with a concentration coefficient of $\alpha=0.5$. 
For MNIST and Fashion-MNIST, 
we employ a ConvNet~\cite{lecun1998gradient}, 
which consists of two convolutional layers and two fully connected layers. 
For CIFAR-10, we utilize VGG11~\cite{simonyan2014very}, 
which comprises eight convolutional layers and three fully connected layers. 
In our experiments, 
for a 4-layer ConvNet, 
all clients share the first three layers. 
Similarly, for VGG11, 
all clients share the first six layers. 

\subsubsection{Training Details} 
We perform 200, 400, and 1000 communication rounds for MNIST, Fashion-MNIST, and CIFAR-10, respectively. 
Across all datasets, 
we set 
the number of local iterations ($\tau$) to 20, 
the local minibatch size ($B$) to 64, 
and the learning rate ($\eta$) to 0.1 with a learning rate decay of 0.99 every 10 rounds. 
We use the SGD optimizer for local training with a weight decay of 1e-4. 
In each round, 10\% of the clients (5 clients) are selected for training and aggregation. 

\subsubsection{Baselines} 
We compare \algo~ with the following baselines: 
\begin{itemize}
  \item Black-box Attack~\cite{qin2023revisiting}: This attack poisons the training data and updates both the feature encoder and classifier simultaneously to minimize the classification loss on both clean and poisoned samples, following standard local training without modification. 
  \item Scaling Attack~\cite{bagdasaryan2020backdoor}: This attack scales its model updates before synchronization with the server. 
  We set the scale factor to 5 in our experiments.
  \item DBA~\cite{xie2020dba}: This attack decomposes the global trigger into multiple parts and distributes them among multiple malicious clients. 
\end{itemize}

\subsubsection{Attack Setup} 
We employ the grid pattern as the trigger, which has been extensively utilized in federated learning backdoor attacks~\cite{xie2020dba,gu2019badnets}. 
Figure~\ref{fig:triggers} illustrates the poisoned images with the grid pattern as the trigger. 
We adopt an all-to-one attack strategy, where all poisoned samples are assigned the same target label regardless of their original ground-truth labels. 
For all datasets, the target label is set to class 2. 
We randomly designate 2 clients as malicious clients. 
Following the attack setting in~\cite{xie2020dba}, 
the malicious clients begin to attack when the main task accuracy converges, 
which is round 50 for MNIST, round 200 for Fashion-MNIST and round 500 for CIFAR-10. 
Subsequently, the malicious clients actively participate in the training process and launch backdoor attacks in every round. 
Following~\cite{zhang2022flip}, in each local iteration, the malicious client poisons 20 samples of a batch on MNIST and Fashion-MNIST, and 5 samples on CIFAR-10. 
For \algo~, we search for the best $\sigma$ values from \{0.01, 0.05, 0.1, 0.15, 0.2\}. 
Without explicit declaration, we set $\sigma$ to 0.2 for MNIST and Fashion-MNIST, and 0.01 for CIFAR-10 in all experiments. 

\subsubsection{Evaluation Details} 
We employ two commonly used metrics, namely, 
the attack success rate (ASR) 
and the main-task accuracy (MTA), 
to evaluate the attack performance of backdoor attacks~\cite{xie2020dba}. 
ASR and MTA denote the classification accuracy of the targeted model on the poisoned and clean testing data, respectively. 
A successful attack should achieve a high ASR while maintaining a high MTA, 
indicating effective manipulation of the model's outputs 
without compromising its performance on the main task. 
We assess the model's performance every two rounds for MNIST, every four rounds for Fashion-MNIST, and every ten rounds for CIFAR-10, respectively. 
Each experiment is executed five times with different random seeds. 
In each run, we calculate the average results over the final 50 rounds. 
For each experiment, we present the average results from these five runs. 

\begin{figure*}[htbp!]
    \centering
    \includegraphics[width=1\linewidth,trim=0 0 0 0,clip]{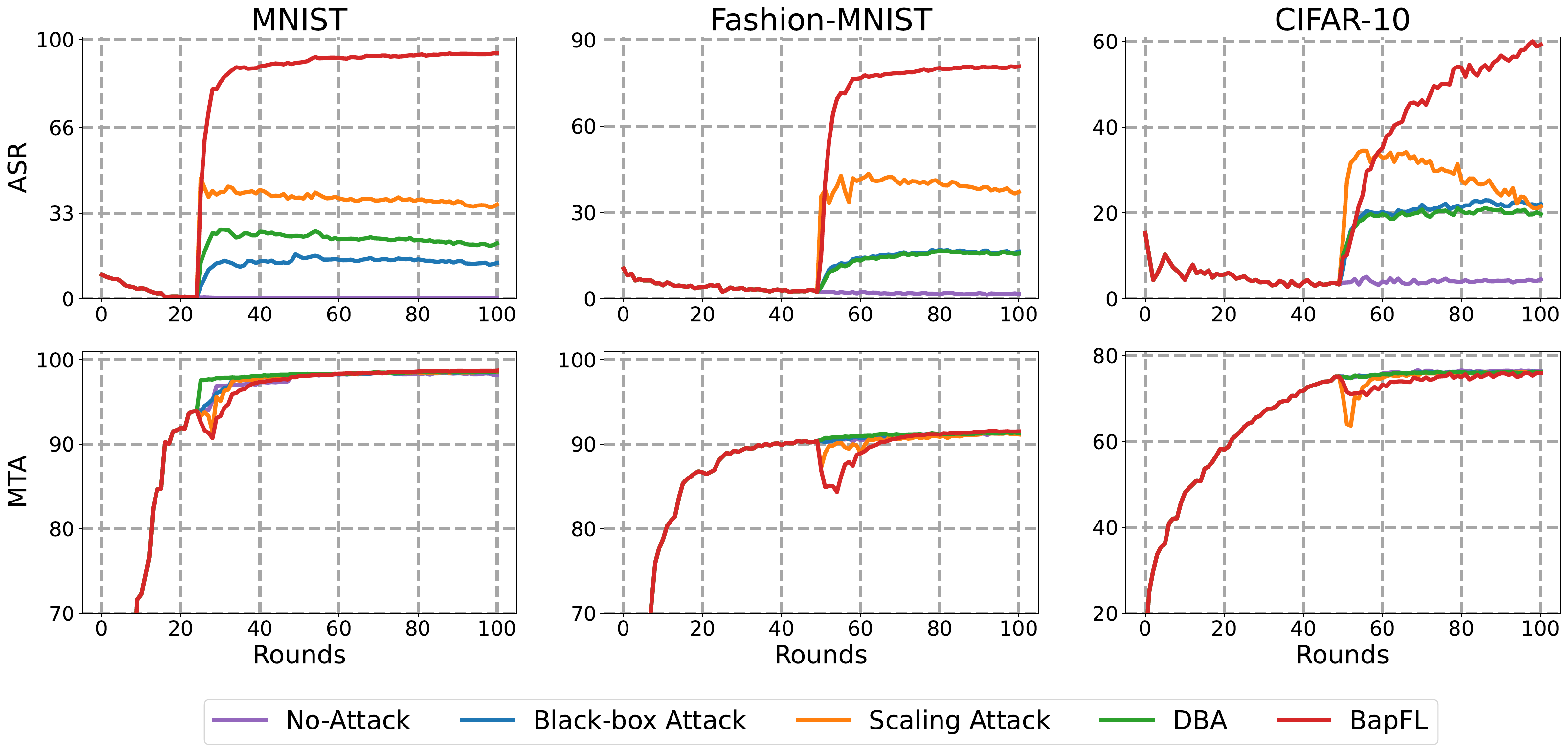}
    \caption{ASR and MTA curves of all attacks on three datasets.}
    \label{fig:overall-performance}
\end{figure*}

\subsection{Main Results}
\label{sec:rq1}
We present the ASR and MTA curves of all attacks in 
Figure~\ref{fig:overall-performance}. 
It is evident that 
\algo~ significantly outperforms other attacks across all datasets 
in terms of attack performance, 
without negatively impacting the model's performance on the main task. 
To be specific, \algo~ surpasses the best baseline, Scaling Attack, by about 57\% on MNIST, 42\% on Fashion-MNIST and 36\% on CIFAR-10. 
As the training progresses, 
there is an increasing divergence in local classifiers between the malicious and benign clients, 
leading to a reduction in ASR for baseline attacks, particularly the Scaling Attack. In contrast, \algo~ demonstrates stable attack performance even in the later stages of training.

\subsection{Data Heterogeneity}
\label{sec:rq2}
\begin{figure*}[htbp!]
    \centering
    \includegraphics[width=1\linewidth,trim=0 0 0 0,clip]{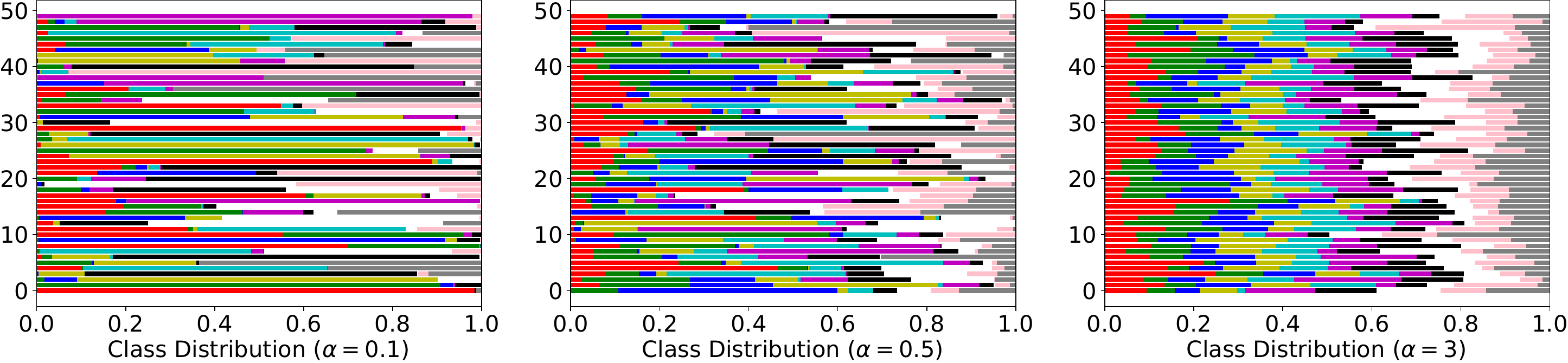}
    \caption{Client class distribution, partitioned by Dirichlet allocation with varying concentration coefficient $\alpha \in \{0.1, 0.3, 0.5, 1, 3\}$.}
    \label{fig:class-distribution}
\end{figure*}
The parameter-decoupling based pFL methods achieve personalization through individualized classifiers. 
And the stronger the data heterogeneity, the larger the divergence in local classifiers among clients. 
The increasing divergence makes it challenging for malicious clients to successfully attack other clients. 
To investigate the impact of data heterogeneity on attack effectiveness, we run \algo~ and the baselines on three datasets. 
We control the degree of data heterogeneity by varying the parameter $\alpha \in \{0.1, 0.3, 0.5, 1, 3\}$. 
The client class distribution is displayed in Figure~\ref{fig:class-distribution}. 
The smaller the value of $\alpha$, the stronger the data heterogeneity. 
Note that when $\alpha \leq 0.5$ , each client is likely to have two or three dominating classes while owning a few or even zero samples from the remaining classes.

\begin{figure*}[htbp!]
    \centering
    \includegraphics[width=1\linewidth,trim=0 0 0 0,clip]{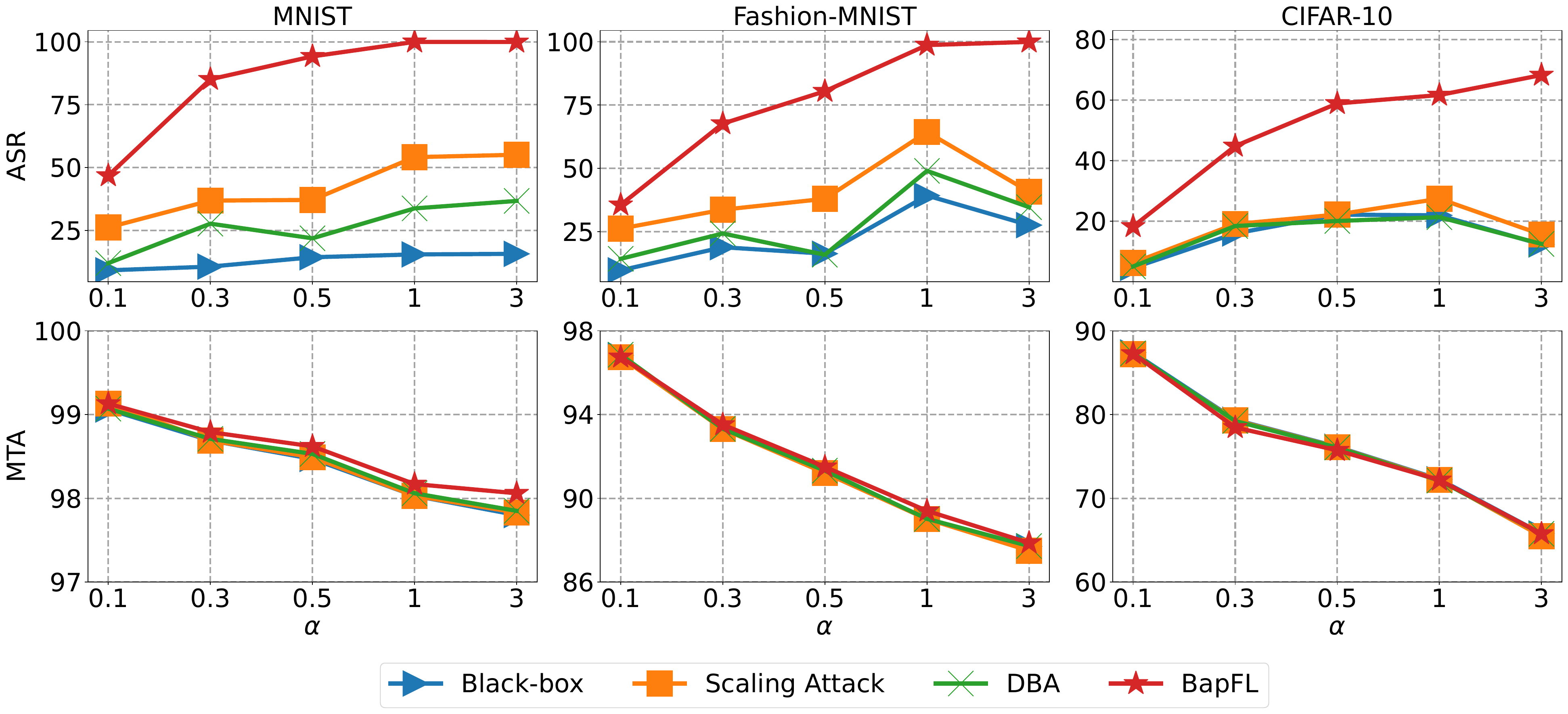}
    \caption{ASR and MTA curves of all attacks on three datasets.}
    \label{fig:varying-alpha}
\end{figure*}
We present the experimental results in Figure~\ref{fig:varying-alpha}. 
It is evident that, with comparable main task performance, \algo~ consistently outperforms the baselines on all datasets across varying $\alpha$. 
In general, for all attacks, as data heterogeneity increases, the ASR decreases. 
However, even under settings with highly strong data heterogeneity ($\alpha$=0.1), \algo~ still achieves an ASR of 46.82\% on MNIST, 35.58\% on Fashion-MNIST and 18.25\% on CIFAR-10, respectively (surpassing the best baseline, Scaling Attack, by about 20\%, 10\% and 12\%, respectively), which still poses a threat to the security of FL systems. 
When $\alpha$=3, the data distribution among clients is relatively similar, and \algo~ achieves an ASR of 99.96\% on MNIST, 99.93\% on Fashion-MNIST, and 68.27\% on CIFAR-10, respectively. 
In contrast, the best baseline, Scaling Attack, achieves an ASR of only 55.10\% on MNIST, 40.77\% on Fashion-MNIST, and 15.59\% on CIFAR-10, respectively. 
Furthermore, \algo~ exhibits greater stability. 
As $\alpha$ increases, the ASRs of \algo~ steadily improves. 
In contrast, the performance of the baselines experiences some degree of fluctuation.

\subsection{Number of Sharing Layers}
\label{sec:rq3}
\begin{figure*}[htbp!]
    \centering
    \includegraphics[width=0.7\linewidth,trim=0 0 0 0,clip]{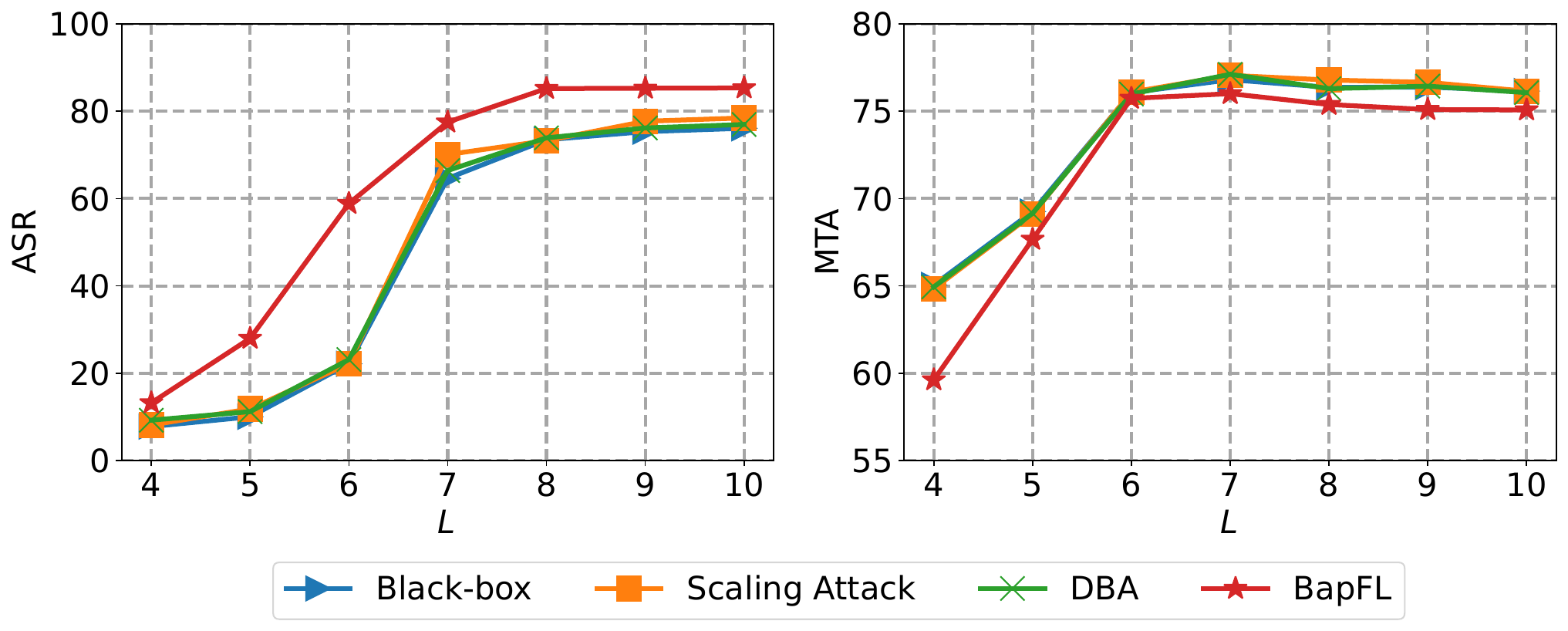}
    \caption{Comparison of \algo~ and the baselines with varying $L$, i.e., the number of sharing layers.}
    \label{fig:varying-L}
\end{figure*}
In this section, 
we investigate the impact of the number of sharing layers, denoted as $L$, on ASR and MTA. 
We conduct experiments with \algo~ and the baselines on CIFAR-10 
with varying $L \in \{4, 5, 6, 7, 8, 9, 10\}$. 
The results in Figure~\ref{fig:varying-L} reveal a significant trade-off: sharing more layers substantially exposes the model to backdoor attacks, while sharing too few layers acts as a defense, albeit with a significant degradation in model utility.
Balancing the trade-off between model utility and security by varying $L$ is a critical challenge. 
We can see that $L=6$ seems a promising choice. 
Compared to $L=7$, 
it can reduce the ASR for the baseline attacks by over 40\% 
with a minimal cost of approximately 1\% in terms of MTA. 
However, even when $L=6$, 
\algo~ can still achieve an ASR of 58.89\%, 
posing a significant threat to the security of pFL systems.

\begin{figure*}[b!]
    \centering
    \includegraphics[width=1\linewidth,trim=0 0 0 0,clip]{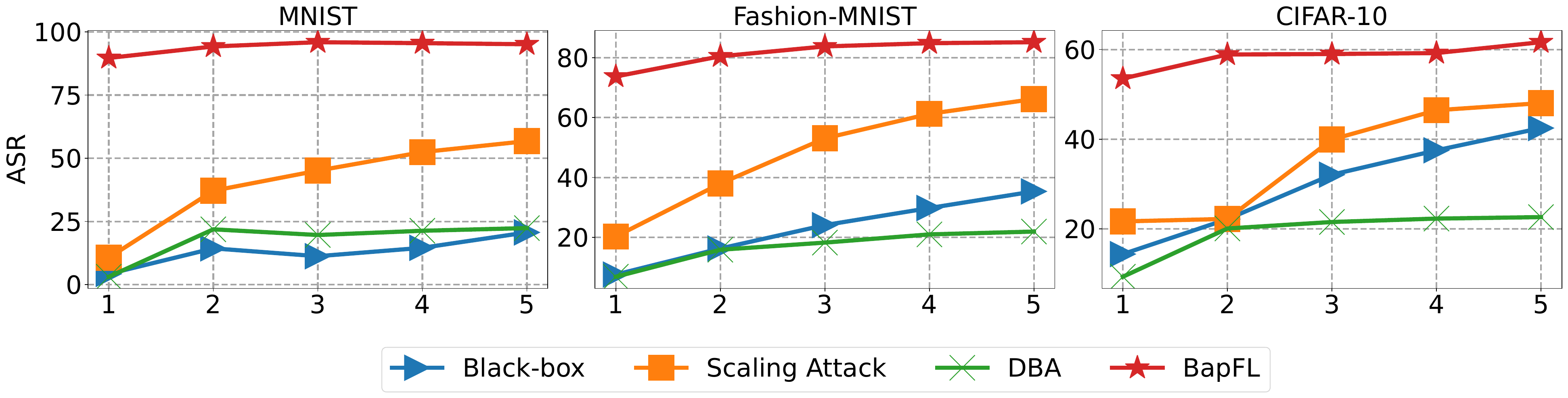}
    \caption{Comparison of \algo~ and the baselines with varying $A$, i.e., the number of malicious clients.}
    \label{fig:varying-attackers}
\end{figure*}
\subsection{Number of Attackers}
\label{sec:rq4}
Intuitively, the larger the number of malicious clients, the broader the attack surface becomes, thus increasing vulnerability for benign clients. 
A straightforward question arises that does \algo~ still pose a threat when there is only one malicious client? 
To investigate this, we conduct experiments with \algo~ and the baselines on three datasets with varying number of malicious clients $A \in \{1, 2, 3, 4, 5\}$. 
For cases where $A \geq 2$, we set the number of participating malicious clients each round to 2, and for $A=1$, we set the number of participating malicious clients each round to 1. The experimental results are presented in Figure~\ref{fig:varying-attackers}. 

It is noteworthy that the baseline attackers require a larger number of malicious clients to enhance their attack performance. 
Even with a total of five attackers, 
Scaling Attack achieves an ASR of only 
56.78\% on MNIST, 
66.20\% on Fashion-MNIST, 
and 48.04\% on CIFAR-10, respectively. 
In contrast, with just one attacker, \algo~ attains an ASR of 
89.67\% on MNIST, 
73.72\% on Fashion-MNIST, 
and 53.52\% on CIFAR-10, 
showcasing its robust generalization capability.

\subsection{Attack Frenquency}
\label{sec:rq5}
As launching attacks at every round may lack stealth, we closely examine the performance of \algo~ with a larger attack interval, such as 16 rounds. 
Additionally, to enhance attack efficiency, we integrate the Scaling Attack to \algo~, denoted as \algo~+Scaling. We conduct experiments with \algo~ and \algo~+Scaling on three datasets with varying attack interval $I \in \{1, 2, 4, 8, 16\}$.

\begin{figure*}[h!]
    \centering
    \includegraphics[width=1\linewidth,trim=0 0 0 0,clip]{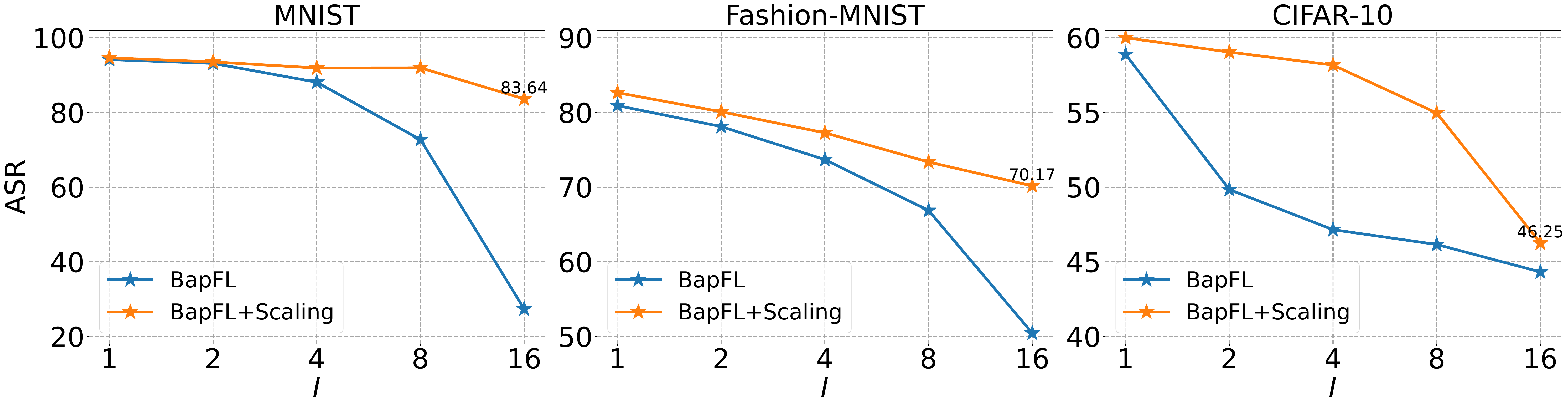}
    \caption{Attack performance of \algo~ and \algo~+Scaling with varying $I$, i.e., the attack interval.}
    \label{fig:attack-interval}
\end{figure*}

The results presented in Figure~\ref{fig:attack-interval} reveal that \algo~ remains effective with $I \leq 8$, but when $I = 16$, \algo~ experiences a significant performance drop. 
However, it is clear that the Scaling operation improves the sustainability of \algo~ by a large margin. 
\algo~+Scaling maintains an ASR of 83.64\% on MNIST, 70.17\% on Fashion-MNIST, and 46.25\% on CIFAR-10 even when $A=16$, respectively. 

\subsection{Ablation Study} 
\label{sec:rq6}
\begin{figure*}[htbp!]
    \centering
    \includegraphics[width=1\linewidth,trim=0 0 0 0,clip]{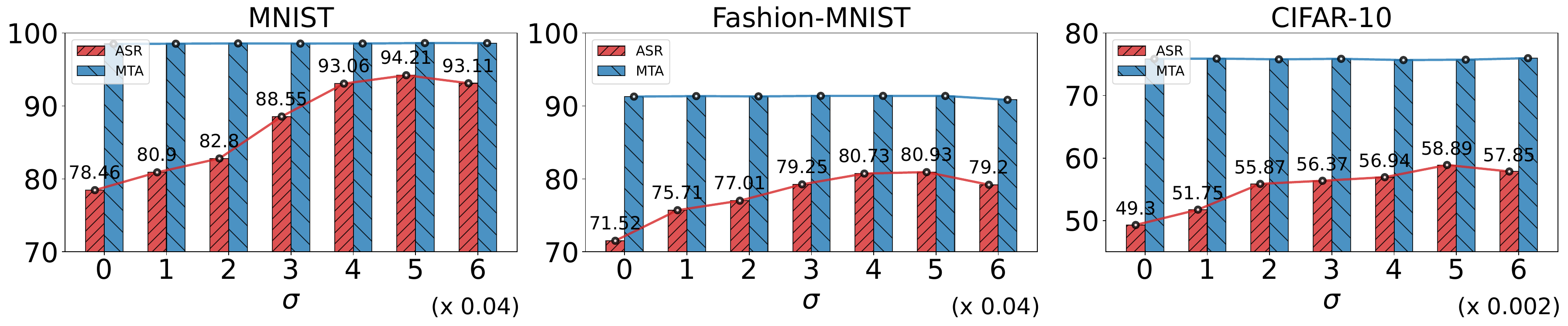}
    \caption{ASR and MTA of \algo~ with varying $\sigma$ on three datasets.}
    \label{fig:varying-K}
\end{figure*}

To generalize the attack to benign clients, 
\algo~ perturbs the parameters of the classifier with noise from a Gaussian distribution $\mathcal{N}(0, \sigma)$. 
When $\sigma=0$, 
\algo~ degrades to its base version, 
denoted as \basealgo~, 
which incorporates only strategy $\text{PS}_{1}$. 
We conduct experiments on three datasets with varying $\sigma$. 
Specifically, we test \algo~ with $\sigma \in 0.04 \times \{0, 1, 2, 3, 4, 5, 6\}$ on MNIST and Fashion-MNIST, and $\sigma \in 0.002 \times \{0, 1, 2, 3, 4, 5, 6\}$ on CIFAR-10. 
The experimental results are presented in Figure~\ref{fig:varying-K}. 

It is evident that increasing $\sigma$ consistently improves the ASR without deteriorating the MTA when $\sigma \leq 0.2$ on MNIST and Fashion-MNIST, and $\sigma \leq 0.01$ on CIFAR-10. 
This can be attributed to that significant noise helps in generalizing attacks to diverse personalized classifiers. 
However, when $\sigma$ exceeds the threshold, the attack performance begins to decline. 
This occurs due to two primary reasons: excessive noise might make the optimization unstable, and the noised classifier is likely to deviate from that of benign clients.
Furthermore, we observe that: 
\begin{itemize}
  \item \basealgo~ achieves an ASR of 78.46\% on MNIST, 71.52\% on Fashion-MNIST, and 49.3\% on CIFAR-10, respectively. These results still outperform the best baseline, Scaling Attack, by about 41\%, 34\%, and 27\%, respectively, thereby validating the effectiveness of strategy $\text{PS}_{1}$. 
  \item \algo~ with the optimal $\sigma$ surpasses \basealgo~ by approximately 15\% on MNIST and 9\% on Fashion-MNIST and CIFAR-10, further validating the effectiveness of strategy $\text{PS}_{2}$. 
\end{itemize}

\section{Defense}
In this section, 
we assess potential defenses 
against \algo~ and the baseline attacks. 
We consider six widely used defense strategies: 
\begin{itemize}
  \item Gradient Norm-Clipping~\cite{sun2019can}: The server normalizes clients' model updates that exceed a predefined threshold of $H$. 
  \item Median~\cite{yin2018byzantine}: The server aggregates clients' updates by calculating their element-wise median. 
  \item Trimmed Mean~\cite{yin2018byzantine}: The server averages clients' updates after excluding a certain percentage ($\beta$) of extreme values from both ends of the parameter distribution. 
  \item Multi-Krum~\cite{blanchard2017machine}: The server calculates the Krum distance for each update 
  by summing the parameter distance between the update and the other closest $M-f$ clients, 
  and then aggregates the updates from $J$ clients with the lowest Krum distance to obtain the final result. 
  Here, $M$ denotes the total number of clients participating in each round, 
  and $f$ is equal or larger than the number of malicious clients. 
  \item Fine-Tuning~\cite{qin2023revisiting}: Clients fine-tune the entire model using their local clean datasets. 
  \item Simple-Tuning~\cite{qin2023revisiting}: Clients reinitialize their classifiers and then retrain them using their local clean datasets while keeping the feature encoder fixed. 
\end{itemize}
We evaluate the performance of these defenses on MNIST with $\alpha=0.5$, as all attacks exhibit the highest ASR among the three datasets. 
Specifically, we assess Gradient Norm-Clipping with different values of $H \in \{0.1, 0.3, 0.5\}$. 
For Trimmed Mean, we select $\beta \in \{0.2, 0.4\}$ to exclude two or four clients from the aggregation process. 
Regarding Multi-Krum, 
the hyperparameter $f$ is set to 2, 
corresponding to the number of malicious clients, 
while $J$ is set to 3, 
corresponding to the number of benign clients. 
For Fine-Tuning, we perform 20 iterations of fine-tuning on the entire model, 
whereas for Simple-Tuning, 
we retrain the local classifier for 200 iterations. 
As the defenses, 
including Gradient Norm-Clipping, 
Median, 
Trimmed Mean, and Multi-Krum, 
tend to slow down the model convergence speed, 
we conduct an additional 200 communication rounds. 
The rest of the experimental setup remains consistent with the details outlined in Section~\ref{sec:exp-setup}. 

Furthermore, 
to bypass Multi-Krum, 
we incorporate PGD~\cite{wang2020attack} into \algo~. 
Specifically, 
let $\omega^{\circ}$ represent the updated classifier on clean samples (line 14 in Algorithm~\ref{alg:algorithm}), and $\omega^{'}$ denote the updated classifier on poisoned samples (line 18 in Algorithm~\ref{alg:algorithm}). 
We project $\omega^{'}$ onto a ball centered around $\omega^{\circ}$ with a radius of $\delta$, i.e., $\Vert \omega^{'} - \omega^{\circ} \Vert \leq \delta$. 
In our experiments, we set $\delta$ to 0.01.

\subsection{Results}
We present the experimental results in Tab.~\ref{tab:defense-results}. 
From the attacker's perspective, 
compared with the baseline attacks, 
\algo~ consistently demonstrates the best attack performance in the presence of all defenses. 
From the defender's perspective, 
among the defenses, 
Multi-Krum offers the most effective protection. 

Specifically, 
it can be observed that 
Gradient Norm-Clipping can reduce \algo~'s ASR from 94.21\% to 46.10\%, but it comes at a cost of approximately 3\% in terms of MTA. 
In contrast, Median and Trimmed Mean with $\beta=0.4$ provide similar defense effects without compromising the model's utility. 
Unfortunately, Fine-Tuning fails to provide effective defense against backdoor attacks and even exacerbates the ASR. 
For instance, it increases the ASR from 21.89\% to 23.61\% for DBA and from 94.21\% to 94.63\% for \algo~. 
Although Simple-Tuning shows improved defense performance compared to Fine-Tuning, 
the ASR of \algo~ still exceeds 85\%. 
Multi-Krum demonstrates superior defense performance, reducing the ASR baseline attacks to below 1\% and that of \algo~ to below 10\%. 
However, when combined with PGD, \algo~ can easily bypass Multi-Krum and obtain an ASR of 70.60\%, indicating a significant threat. 

\begin{table*}[t]
\caption{
Performance comparison among existing defenses. 
An effective defense is characterized by a lower ASR ($\downarrow$) and a higher MTA ($\uparrow$). 
The best attack results are highlighted in \textbf{bold}, while the most effective defenses are marked with $\dagger$. 
} 
\resizebox{\linewidth}{!}{

\begin{tabular}{|c|cc|cccccc|cc|cccc|cc|cc|cc|}
\hline
\multirow{3}{*}{Defenses} 
& \multicolumn{2}{c|}{No-Defense}       
& \multicolumn{6}{c|}{Gradient Norm-Clipping} 
& \multicolumn{2}{c|}{Median}
& \multicolumn{4}{c|}{Trimmed Mean} 
& \multicolumn{2}{c|}{Multi-Krum}   
& \multicolumn{2}{c|}{Fine-Tuning}      
& \multicolumn{2}{c|}{Simple-Tuning} 
\\ 
\cline{2-21} 
& \multicolumn{2}{c|}{\textbackslash{}} 

& \multicolumn{2}{c|}{$H=0.1$}                              
& \multicolumn{2}{c|}{$H=0.3$}                              
& \multicolumn{2}{c|}{$H=0.5$}        

& \multicolumn{2}{c|}{\textbackslash{}} 

& \multicolumn{2}{c|}{$\beta=0.2$} 
& \multicolumn{2}{c|}{$\beta=0.4$} 

& \multicolumn{2}{c|}{\textbackslash{}}

& \multicolumn{2}{c|}{\textbackslash{}}

& \multicolumn{2}{c|}{\textbackslash{}}

\\ 
\cline{2-21} 
                          
& ASR & MTA    
& ASR & \multicolumn{1}{c|}{MTA} & ASR & \multicolumn{1}{c|}{MTA} & ASR & MTA 
& ASR & MTA 
& ASR & \multicolumn{1}{c|}{MTA} & ASR & MTA 
& ASR & MTA 
& ASR & MTA 
& ASR & MTA 
\\ 
\hline
Black-box Attack          
& 14.36 & \multicolumn{1}{c|}{98.47} 
& \textcolor{white}{0}6.60 & \multicolumn{1}{c|}{96.63} 
& 10.99 & \multicolumn{1}{c|}{97.60} 
& 12.05 & \multicolumn{1}{c|}{97.95} 
& \textcolor{white}{0}6.74 & \multicolumn{1}{c|}{98.40} 
& \textcolor{white}{0}9.47 & \multicolumn{1}{c|}{98.34} 
& \textcolor{white}{0}6.87 & \multicolumn{1}{c|}{98.38} 
& \textcolor{white}{0}0.47$\dagger$ & \multicolumn{1}{c|}{98.50} 
& 11.80 & \multicolumn{1}{c|}{98.51} 
& \textcolor{white}{0}5.05 & \multicolumn{1}{c|}{98.22} 
\\ 
Scaling Attack          
& 37.15 & \multicolumn{1}{c|}{98.49}
& 15.14 & \multicolumn{1}{c|}{96.95}
& 19.66 & \multicolumn{1}{c|}{97.72} 
& 20.52 & \multicolumn{1}{c|}{98.02} 
& \textcolor{white}{0}8.35 & \multicolumn{1}{c|}{98.46} 
& 15.83 & \multicolumn{1}{c|}{98.36} 
& \textcolor{white}{0}7.85 & \multicolumn{1}{c|}{98.43} 
& \textcolor{white}{0}0.40$\dagger$ & \multicolumn{1}{c|}{98.48}
& 26.89 & \multicolumn{1}{c|}{98.52} 
& 17.94 & \multicolumn{1}{c|}{98.13} 
\\ 
DBA          
& 21.89 & \multicolumn{1}{c|}{98.53}
& 6.08  & \multicolumn{1}{c|}{96.62} 
& 13.70 & \multicolumn{1}{c|}{97.66}
& 15.77 & \multicolumn{1}{c|}{97.97} 
& \textcolor{white}{0}7.85 & \multicolumn{1}{c|}{98.41} 
& 14.14 & \multicolumn{1}{c|}{98.33} 
& \textcolor{white}{0}8.02 & \multicolumn{1}{c|}{98.41} 
& \textcolor{white}{0}0.33$\dagger$ & \multicolumn{1}{c|}{98.41}
& 23.16 & \multicolumn{1}{c|}{98.55} 
& 10.62 & \multicolumn{1}{c|}{98.24} 
\\ 
\algo~          
& \textbf{94.21} & \multicolumn{1}{c|}{98.62}
& \textbf{46.10} & \multicolumn{1}{c|}{95.71}
& \textbf{80.24} & \multicolumn{1}{c|}{97.46} 
& \textbf{88.52} & \multicolumn{1}{c|}{97.95} 
& \textbf{46.79} & \multicolumn{1}{c|}{98.45} 
& \textbf{94.20} & \multicolumn{1}{c|}{98.61} 
& \textbf{52.91} & \multicolumn{1}{c|}{98.52} 
& \textbf{\textcolor{white}{0}9.18}$\dagger$ $\xrightarrow{\text{PGD}}$ \textbf{70.60} & \multicolumn{1}{c|}{97.09 $\xrightarrow{\text{PGD}}$ 97.11}
& \textbf{94.63} & \multicolumn{1}{c|}{98.69} 
& \textbf{88.65} & \multicolumn{1}{c|}{98.08} 
\\  
\bottomrule
\end{tabular}

}
\label{tab:defense-results}
\end{table*}

\section{Conclusion}
In this paper, 
we have shed light on the vulnerability of 
personalized federated learning (pFL) methods 
with parameter decoupling to backdoor attacks, emphasizing the necessity for robust defenses. 
Initially, we analyzed two fundamental factors contributing to the inherent challenges in attacking personalized models. 
Then, we introduced two straightforward poisoning strategies individually to counteract the two factors. 
Combining these two strategies, we developed a practical and easy-to-implement backdoor attack, namely \algo~. 
Extensive experiments have demonstrated the effectiveness of \algo~. 
Furthermore, we assessed the efficacy of six widely used defense methods and found that \algo~ still remains a significant threat even in the presence of the best defense, Multi-Krum. 
We hope that our work can stimulate further investigations into both attack and defense strategies within the pFL scenarios, ultimately enhancing the security and trustworthiness of pFL in real-world applications.

\begin{acks}
This work was supported by the National Natural Science Foundation of China under grant number 62202170 and the Ant Group.
\end{acks}

%%
%% The next two lines define the bibliography style to be used, and
%% the bibliography file.
\bibliographystyle{ACM-Reference-Format}
\bibliography{sample-base}

%%
%% If your work has an appendix, this is the place to put it.
\appendix

\end{document}